\documentstyle[aps,twocolumn,psfig,prb,floats]{revtex}

\begin{document}

\draft

\twocolumn[\hsize\textwidth\columnwidth\hsize\csname
@twocolumnfalse\endcsname

\title{Mixed Lattice and Electronic States in High-Temperature Superconductors}

\author{R. J. McQueeney\thanks{email: mcqueeney@lanl.gov}, J. L.
Sarrao, P. G. Pagliuso}
\address{Los Alamos National Laboratory, Los Alamos, New Mexico 87545}

\author{P. W. Stephens}
\address{Department of Physics and Astronomy, State University of New York at Stony Brook, Stony Brook, NY 11794 USA}

\author{R. Osborn}
\address{Argonne National Laboratory, Argonne, Illinois 60439}
\date{Received on April 6, 2001}

\maketitle

\begin{abstract}
Inelastic neutron scattering measurements are presented which show
the abrupt development of new oxygen lattice vibrations near the
doping-induced metal-insulator transition in
La$_{2-x}$Sr$_{x}$CuO$_{4}$.  A direct correlation is established
between these lattice modes and the electronic susceptibility (as
measured by photoemission) inferring that such modes mix strongly
with charge fluctuations. This electron-lattice coupling can be
characterized as a localized one-dimensional response of the
lattice to short-ranged metallic charge fluctuations.

\end{abstract}

]

High-temperature superconductors are based on antiferromagnetic
insulating materials caused by strong electronic correlations.
Doping charge carriers (holes) into this system creates a
two-dimensional correlated metallic state in the CuO$_{2}$ plane
that becomes superconducting at low temperatures.\cite{kastner}\
Much effort has been focused on the role of antiferromagnetic spin
fluctuations in transport properties and superconductivity in the
metallic state, mainly because it is believed that superconducting
transition temperatures are too high to arise solely from
electron-phonon coupling. Recently, this correlated metallic state
has been discussed in terms of localized, atomic-scale charge
density fluctuations.\cite{zaanen,tranquada,yamada}\  As hole
doping affects mainly the hybridized Cu 3d$_{x^{2}-y^{2}}$ and O
2p$_{(x,y)}$ anti-bonding electronic states, the lattice should
couple strongly to a localized matrix of slow charge fluctuations,
perhaps decisively influencing charge transport and
superconducting properties. Inelastic neutron scattering
measurements of the lattice dynamics do show evidence of strong
and unusual electron-lattice coupling in high-T$_{c}$ compounds
with various crystal structures and methods of
doping.\cite{pintschovius91,pintschovius96,reichardt,mcqueeney}\
However, many issues surrounding this anomalous coupling remain
unclear. Of course, the ultimate issue concerns its role in
superconductivity. Here, we report the systematic development of
these anomalous modes on hole concentration and the abrupt
formation of new oxygen lattice vibrations near the
metal-insulator transition (MIT) in La$_{2-x}$Sr$_{x}$CuO$_{4}$.
These new lattice modes may be associated with the bosons which
interact with electronic states, causing a kink in the electronic
dispersions observed by photoemission.\cite{Bogdanov,Lanzara}\
These results are the best experimental evidence yet that the
lattice is strongly mixed with charge dynamics in the
high-temperature superconductors.

\begin{figure}[h]
\psfig{file=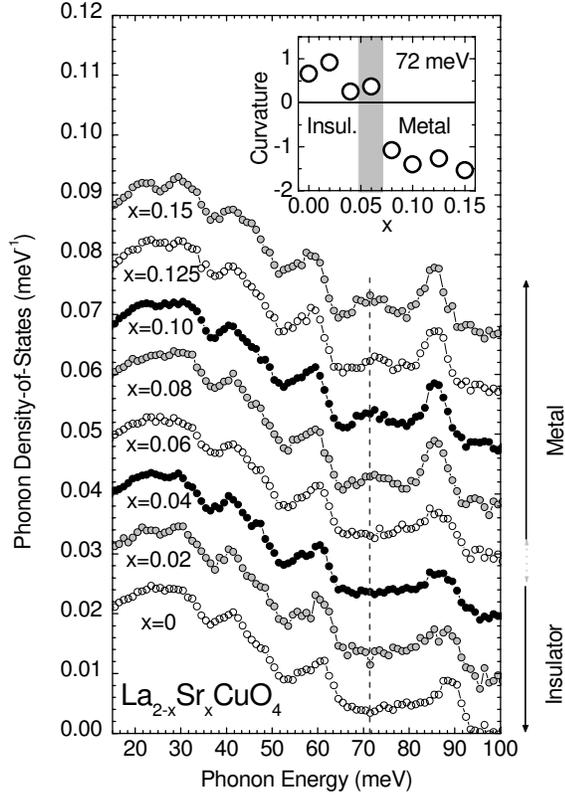,width=.45\textwidth} \caption{The
phonon densities-of-states of La$_{2-x}$Sr$_{x}$CuO$_{4}$ for
several hole concentrations at $T=10~K$. Each DOS is displaced
along the $y$-axis for clarity and the vertical dashed line
indicates the new band formation.  The scale on the right
indicates schematically the metal-insulator transition as a
function of doping.  The inset shows the local curvature of the
DOS at 72 meV as a function of Sr doping.  The curvature was
calculated after smoothing the DOS by convolution with a gaussian
of 2 meV standard deviation.} \label{fig1}
\end{figure}

We performed inelastic neutron scattering measurements of phonon
densities of states (DOS) in La$_{2-x}$Sr$_{x}$CuO$_{4}$ (LSCO)
spanning hole concentrations from the undoped insulator to the
optimally doped superconductor ($0 \leq x \leq 0.15$). The
measurements were performed at $T=10~K$ on the LRMECS spectrometer
at the Intense Pulsed Neutron Source at Argonne National
Laboratory.  Sample preparation details and extraction of the DOS
from the raw data are similar to those previously reported for
La$_{2-x}$Sr$_{x}$NiO$_{4}$.\cite{mcqueeney2}\ The results of the
present measurements are shown in Fig.~\ref{fig1}.  Of primary
importance is the abrupt development of new lattice modes near
$\sim$70 meV for hole concentrations between 6-8\%
($x=0.06-0.08$). Because the DOS is not an analytic function and
cannot be reliably fit to a series of peaks, this development is
characterized by calculating the curvature of the DOS at 72 meV
(inset of Fig.~\ref{fig1}) which changes sign in this
concentration range.

The abrupt change in the DOS is not due to the
tetragonal-orthorhombic structural phase transition, as all
samples remain in orthorhombic at $T=10~K$.\@\cite{radaelli}\  The
new band is also not caused by electrostatic impurity effects from
Sr substitution because identical measurements on the
isostructural nickelate compounds show no such band formation at
these Sr concentrations.\cite{mcqueeney2}\  Rather, the abrupt
development of the 70 meV band must be related to the
doping-induced MIT near $x\sim0.05$, where trapped holes begin to
become mobile.\cite{kastner,imada}\ The onset of superconductivity
and dynamic incommensurate spin fluctuations (signatures of the
stripe correlations) also occurs in this critical concentration
range. The nature of this MIT is unclear at present, however the
transition does occur without a corresponding change in lattice
symmetry.  Thus, the 70 meV band signifies strong and unusual
electron-lattice coupling in the metallic state of high
temperature superconductors that is not present in the insulating
state.  The MIT also affects lower energy phonons, especially near
30 meV.\@  Unfortunately, a plethora of phonon modes existing at
this energy make analysis of the low energy DOS features
difficult, and they are not discussed further.

\begin{figure} [h]
\psfig{file=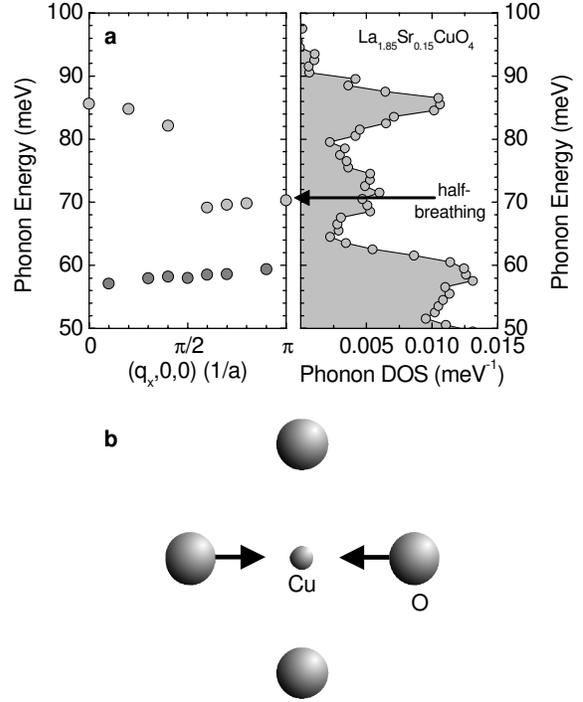,width=.45\textwidth}
\caption{Half-breathing mode in dispersion and DOS measurements.
(a) The high-frequency phonon dispersion of superconducting
La$_{1.85}$Sr$_{0.15}$CuO$_{4}$ along $(1,0,0)$ is shown in the
left panel (from McQueeney {\it et al.} \cite{mcqueeney}).  The
flat mode originating at $(\pi,0)$ and $\sim$70 meV is the
half-breathing oxygen mode.  The lower branch at 60 meV is the
bond-bending branch.  On the right, the high-frequency portion of
the DOS of La$_{1.85}$Sr$_{0.15}$CuO$_{4}$ is shown. The $\sim$70
meV band is comprised at least partly of the anomalous
half-breathing modes. (b) Polarization of the half-breathing
oxygen mode.} \label{fig2}
\end{figure}

The new lattice modes at 70 meV consist, at least partially, of
half-breathing-like oxygen phonon modes that propagate in the
CuO$_{2}$ plane.  Phonon dispersion measurements of LSCO by
inelastic neutron scattering show that the half-breathing oxygen
modes between $\vec{q}=(\pi/2,0,0)$ and $(\pi,0,0)$ soften
anomalously from 80 meV ($x=0$) to 70 meV ($x=0.15$) with doping.
\cite{pintschovius91,mcqueeney}\ Figure~\ref{fig2}(a) demonstrates
clearly that this flat half-breathing branch contributes to the
DOS as a van Hove singularity for La$_{1.85}$Sr$_{0.15}$CuO$_{4}$
near 70 meV.\@ The half-breathing polarization is shown in
Fig.~\ref{fig2}(b).\cite{pintschovius91}\

\begin{figure} [h]
\psfig{file=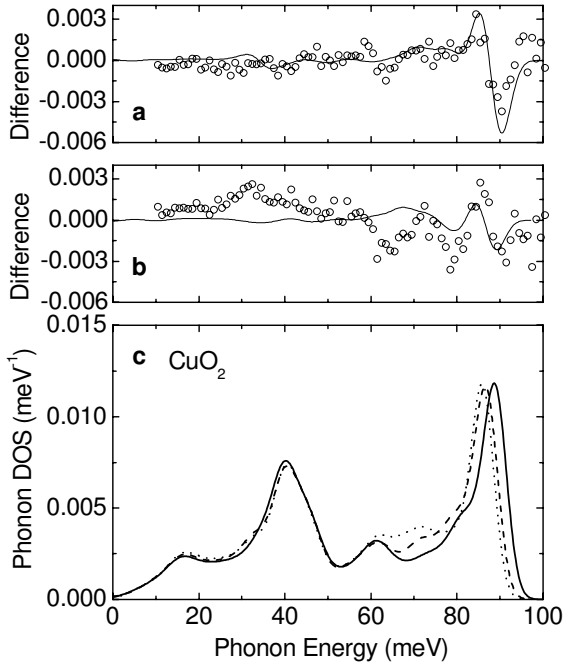,width=.45\textwidth}
\caption{Comparison to two-dimensional DOS calculations. (a) The
difference of the two model DOS calculations for $x=0.06$ and
$x=0$ (line) and the difference in the data (circle).  (b) The
difference of the two model DOS calculations for $x=0.08$ and
$x=0.06$ (line) and the difference in the data (circle).  (c)  The
calculated CuO$_{2}$ phonon density-of-states for $x=0$ (full
line), $x=0.06$ (dashed), and $x=0.08$ (dotted).} \label{fig3}
\end{figure}

To learn more about the nature of the electron-lattice coupling of
the half-breathing modes, we use a simple lattice dynamics model
to isolate the effective interaction being mediated by the mobile
holes.  We are only interested in the highest energy modes
propagating within CuO$_{2}$ plane.  Also, for these vibrational
energies (above 60 meV) the phonons are comprised entirely of
in-plane polarized oxygen modes.  Therefore, a two-dimensional
harmonic force constant (ball and spring) model of the CuO$_{2}$
plane is used to reproduce the high-frequency DOS. The initial set
of pairwise force constants are chosen to reproduce the measured
in-plane phonon dispersion and DOS of undoped La$_{2}$CuO$_{4}$.
Doping is introduced by varying the magnitude of the force
constants while maintaining the full periodicity of the CuO$_{2}$
plane. Models studied with spatially varying force constants
(superlattice models) produced only small variations in the DOS.
The model reproduces well the subtle changes observed in the high
frequency oxygen phonons within the insulating phase ($x<0.06$),
such as the gradual weak softening of the $\sim$88 meV phonon
band, by reducing the Cu-O nearest-neighbor force constant. This
is shown in figure~\ref{fig3}(a).

The model has difficulty reproducing the large changes of the
oxygen phonons in the metallic state at $x=0.08$. One is required
to introduce a repulsive force constant between next-nearest
neighbor oxygens with a bridging copper to produce a band near 70
meV originating from the half-breathing mode.  More importantly,
this repulsion must be introduced anisotropically between only one
oxygen pair in a given CuO$_{4}$ plaquette (not the orthogonal
pair), thereby breaking the crystal symmetry (although perhaps
only locally and over phonon time scales). Introducing the
repulsion over both pairs strongly softens oxygen breathing modes
(near 85 meV) in addition to the half-breathing modes,
inconsistent with experimental results.  The calculated and
measured differences in the $x=0.06$ and $0.08$ DOS are shown in
figure~\ref{fig3}(b) and results of the full DOS for $x=0$,
$0.06$, and $0.08$ are shown in figure~\ref{fig3}(c). This simple
model suggests that the electron-lattice coupling is
one-dimensional in nature and the coupled hole states are oriented
along the Cu-O-Cu bond direction.

However, differences between the metallic and insulating DOS
cannot be explained fully with the simple model.  The intensity of
the 70 meV band is relatively large implying that many phonons
($\sim$ 15\% of all possible oxygen modes) are affected.  The
metallic phonon bands are also narrower in energy than the
insulating bands, which suggests phonon localization.  The
sharpness and intensity of the metallic bands are consistent with
phonon dispersion measurements in LSCO where the
half-breathing-like modes are observed to be dispersionless in a
large region of the Brillouin zone between $\pi/2a
<|q_{x}|<\pi/a$, $|q_{y}|<0.3\pi/a$, and independent of
$q_{z}$.\cite{mcqueeney}\ This entire Brillouin zone pocket
centered at $(\pi,0,0)$ forms the rather intense van Hove
singularity in the DOS at 70 meV.\@ These observations cannot
arise from the usual electron-phonon coupling in a metal which
causes Kohn anomalies in the phonon dispersion at specific
wavevectors $\vec{q}=2\vec{k}_{F}$ (where $\vec{k}_{F}$ is the
Fermi surface wavevector).

Strong electron-lattice coupling and the one-dimensionality of the
charge states are corroborated by various angle-resolved
photoemission results.  Bogdanov {\it et al.}\cite{Bogdanov} show
that the electronic band dispersion has a kink due mixing of holes
with other bosonic excitations in the range of oxygen optical
phonon energies. Lanzara {\it et al.} \cite{Lanzara} conclusively
identify these excitations as phonons. For LSCO, Lanzara {\it et
al.} also show that strong electron-lattice coupling occurs with
phonons in the 70 meV energy range, i.e. the half-breathing modes.
Ino {\it et al.} \cite{Ino} have studied the doping dependence of
the photoemission in La$_{2-x}$Sr$_{x}$CuO$_{4}$. These results
reveal that the $(\pi,0,0)$ electronic saddlepoint sits $\sim$500
meV below the Fermi level in the insulating phase.  Doping moves
the saddlepoint close to the Fermi level, eventually producing a
MIT at $x\sim0.05$ where charge dynamics are characterized as
one-dimensional.  The maximum spectral weight of the saddlepoint
band crosses the half-breathing phonon energy (70-80 meV) near
$x=0.07$ consistent with new phonon modes being observed somewhat
above the critical hole concentration of the MIT as measured by
transport. This behavior is unclear, but may arise from the
disparity in the time scales of inelastic neutron scattering and
transport measurements.

It is possible that the one-dimensional charge fluctuations
originate from states in the extended electronic saddlepoint near
$\vec{q}=(\pi,0,0)$. Then, half-breathing-like phonons and
saddlepoint holes form localized states (or "wave packets" made up
of many plane waves). The stripe scenario supports this conjecture
by assuming an inhomogeneous and localized charge distribution in
the metallic state \cite{zaanen,tranquada}. Castro Neto has shown
that strong phonon and electronic responses at $(\pi,0,0)$ occur
in the stripe model from the scattering of hole pairs at stripe
domain walls.\cite{neto}\  Another source of charge inhomogeneity
originates from large Peierls-type (phonon induced) charge
transfer fluctuations. The introduction of charge fluctuations
into electronic calculations within the local-density
approximation is known to produce strong screening of the
half-breathing phonon modes in La$_{2}$CuO$_{4}$.\cite{falter}
Results obtained from exact diagonalization of the one-dimensional
Peierls-Hubbard model also indicate that charge transfers caused
by half-breathing modes are enhanced when strong correlations are
included.\cite{ishihara}\ These results suggest that the phonon
softening in the metallic state is likely due to the strong local
interaction of phonons with charge fluctuations above the
metal-insulator transition. This important interaction bears
consideration in any theory of high temperature superconductivity.

\acknowledgments

RJM would like to thank T. Egami, A. Castro Neto and A. Ramirez
for helpful discussions.  This work was supported by the U.S.
Department of Energy under contract number W-7405-Eng-36 with the
University of California.  This work has benefited from the use of
the Intense Pulsed Neutron Source at Argonne National Laboratory.
This facility is funded by the U. S. Department of Energy,
BES-Materials Science, under contract W-31-109-Eng-38.


\end{document}